\documentclass[asmfonts,superscriptaddress,amssymb,showkeys, nofootinbib,onecolumn,notitlepage,nobibnotes, aps,prd,10pt]{revtex4-1}

\usepackage[english]{babel}
\usepackage[pdftex, pdftitle={Article}, pdfauthor={Author}]{hyperref}
\usepackage[utf8]{inputenc}
\usepackage[T1]{fontenc}
\usepackage[english]{babel}
\usepackage{ifpdf,amsmath,amsthm,amssymb,amsfonts,newtxtext,newtxmath,mathtools} 
\usepackage{calc}
\usepackage{xcolor}
\usepackage{physics}
\usepackage{subcaption} 
\usepackage{amsmath}
\usepackage{comment}

\newcommand{\coeffD}{\beta}

\begin{document}

\title{Geometric entanglement in integer quantum Hall states with boundaries}

\author{Pierre-Gabriel Rozon}  
\address{D\'epartement de Physique, Universit\'e de Montr\'eal, Montr\'eal, Qu\'ebec, H3C 3J7, Canada}
\author{Pierre-Alexandre Bolteau}
\address{Unit\'e de Formation et de Recherche de Physique, Sorbonne Universit\'e, 75252 Paris, France}
\author{William Witczak-Krempa}
\email{w.witczak-krempa@umontreal.ca}
\address{D\'epartement de Physique, Universit\'e de Montr\'eal, Montr\'eal, Qu\'ebec, H3C 3J7, Canada}
\address{Centre de Recherches Math\'ematiques, Universit\'e de Montr\'eal; P.O. Box 6128, Centre-ville Station; Montr\'eal (Qu\'ebec), H3C 3J7, Canada}
\address{Regroupement Qu\'eb\'ecois sur les Mat\'eriaux de Pointe (RQMP)}

\date{\today}

\begin{abstract}\vspace{4pt}
\begin{center}\textbf{\abstractname}\end{center}\vspace{-5pt} 
Boundaries constitute a rich playground for quantum many-body systems because they can lead to novel degrees of freedom such as protected boundary states in topological phases.
Here, we study the groundstate of integer quantum Hall systems in the presence of boundaries through the reduced density matrix of a spatial region. We work in the lowest Landau level and choose our region to intersect the boundary at arbitrary angles. The entanglement entropy (EE) contains a logarithmic contribution coming from the chiral edge modes, and matches the corresponding conformal field theory prediction. We uncover an additional contribution due to the boundary corners. 
We characterize the angle-dependence of this boundary corner term, and compare it to the bulk corner EE. We further analyze the spatial structure of entanglement via the eigenstates associated with the reduced density matrix, and construct a spatially-resolved EE. The influence of the physical boundary and the region's geometry on the reduced density matrix is thus clarified. Finally, we discuss the implications of our findings for other topological phases, as well as quantum critical systems such as conformal field theories in 2 spatial dimensions. 
\end{abstract}

\maketitle

\section{Introduction} \label{intro}
Boundaries, or more generally interfaces, lead to novel degrees of freedom. For instance, in gapped topological phases, the existence of a topological invariant that changes across the interface gives rise to robust boundary modes. In quantum Hall states, one has chiral one-dimensional edge modes that can be described by an
effective Conformal Field Theory (CFT)~\cite{WenBook}.
In fact, there is a profound correspondence between the bulk, effectively described by a Chern-Simons topological quantum field theory, and the
CFT living on the boundary~\cite{WenBook}.   
A more microscopic and complete understanding of quantum Hall states can be achieved by studying model wavefunctions, for example the Laughlin states~\cite{Laughlin1983} at filling $\nu=1/m$, where $m$ is an odd integer. If $m>1$, such a wavefunction describes a strongly correlated fractional quantum Hall (FQH) state with anyon excitations. The price to pay with model wavefunctions, however, is that one often needs to work numerically with a relatively small number of electrons. Interestingly, it was observed that the entanglement properties of a quantum Hall groundstate contain detailed information not only about the anyonic properties~\cite{Kitaev2006,Levin2006,Dong2008,FradkinBook}, but also about edge modes~\cite{Li2008}. For example, such information can be numerically extracted using model wavefunctions defined on convenient geometries such as the sphere or torus.       

More precisely, one can extract these properties from the reduced density matrix through its entanglement spectrum and entanglement entropy (EE). Given a partition of the total Hilbert space into a product $\mathcal H=\mathcal H_A\otimes\mathcal H_{A^c}$, the reduced density matrix of a state $\rho$ is $\rho_A=\Tr_{A^c}\rho=\exp(-H_A)$.
The entanglement spectrum is the spectrum of the entanglement Hamiltonian $H_A$, while the von Neumann EE is $S_A=-\Tr_A{\rho_A\ln \rho_A}$.
A useful choice of partition is to divide space into a subregion $A$ and its complement $A^c$. By studying how the reduced density matrix changes as a function of the size, shape or topology of region $A$ one can learn a great deal about a quantum state. In particular, topological quantum field theory tells us that for the groundstate of a quantum Hall state on the plane, the EE of a smooth and simply connected region (like a disk) is~\cite{Kitaev2006,Levin2006,Dong2008,FradkinBook}:
\begin{align} \label{TEE}
    S_A = \beta \frac{P_A}{\ell_c} - \gamma
\end{align}
The first term is the boundary law, where $P_A$ is the perimeter of $A$, while $\ell_c$ is a short-distance cutoff.    
When dealing with electronic wavefunctions, and not their effective quantum field theory, this cutoff will be the magnetic length, $\ell_B$. 
In contrast, $-\gamma$, called the topological entanglement entropy (TEE), is universal\footnote{When dealing with a topological gauge theory such as Chern-Simons, care is needed due to the fact that the Hilbert space does not factorize spatially.}. It does not depend on $\ell_c$ nor on the shape of $A$ and equals $-\ln \mathcal D$, where $\mathcal D$ is the total quantum dimension of the topological phase. For a $\nu=1/m$ Laughlin state, $\mathcal D=\sqrt m$, which reveals the existence of anyons at $m\!>\!1$, while it vanishes for the integer quantum Hall state. 
If we introduce a physical boundary or interface, an interesting and natural possibility is to have region $A$ touch the boundary. The EE should then receive
additional \emph{geometrical} contributions. 
For one, the boundary of a Hall fluid can be described by a 1 dimensional CFT so one expects the logarithmic correction proportional to the Virasoro central charge $c$. This was indeed numerically observed in the study of specific FQH wavefunctions~\cite{Varjas2013,crepel_2019,crepel_2019_2,crepel_microscopic_2019}. An additional correction will come due to the presence of boundary corners, i.e.\ intersections between the boundary of $\partial A$ (called the entangling surface) and the physical boundary. These intersections
will induce a corner dependent term to the entropy for each corner, 
$-b(\theta)$, where $\theta$ is the corner's angle. The information encoded in such boundary corners is currently poorly understood, and has scarcely been studied in topological systems. Boundary corners have been recently studied for the groundstates of quantum critical Hamiltonians such as CFTs in 2 spatial dimensions~\cite{Fradkin2006,Fursaev2016,Seminara2017,Berthiere2019-1,Berthiere2019}. In that case, $b(\theta)$ comes multiplied by a logarithm $\ln(P_A/\ell_c)$. By studying the behavior of $b(\theta)$ near $\pi/2$, it was found that they encode essential information about the CFT, such as conformal anomaly coefficients, and other central charges related to the stress tensor~\cite{Fursaev2016,Seminara2017,Berthiere2019}.
Away from orthogonality, little is understood. In this work, we initiate the geometrical investigation of boundary corner entanglement in topological phases.    

Working with the IQH state at $\nu=1$, we analyze the reduced density matrix associated with a region that intersects the boundary. We impose hard-wall boundary conditions (Dirichlet) on the electronic wavefunctions. We first analyze the EE, and find the expected logarithmic subleading term coming from the chiral edge modes. We extract a central charge $c=1$ in agreement with the CFT prediction. We next study the contribution arising from the boundary corners, $b(\theta)$. We find that it scales the same way as in quantum critical systems when $\theta\to \pi/2$ and $\theta\to 0$. 
We further examine the relation between the boundary corner function $b(\theta)$, and its counterpart for corners in the bulk, $a(\theta)$. In the context of quantum field theories with boundaries, a relation was recently proposed~\cite{Berthiere2019}, which in its simplest form is $a(2\theta)=2b(\theta)$. We find that $a(2\theta)\simeq 2b(\theta)$, with deviations being the largest at small angles. We push our analysis beyond the EE, by examining the entanglement spectrum and eigenfunctions associated with the reduced density matrix. The latter allow us to gain insights into the spatial structure of entanglement, and construct a spatially-resolved EE, $\tilde S_A(\vec r)$. The influence of the physical boundary and the region's geometry, in particular the corners, is thus clarified. We end by discuss ramifications of our results to other topological systems, as well as quantum critical states.   

The paper is organized as follows. In Section~\ref{setup} we describe the Hamiltonian and its spectrum. Section~\ref{Results} is concerned with the geometrical dependence of the EE and entanglement spectrum. In particular, we elucidate the edge mode contribution to the EE, and study the corner contribution $b(\theta)$. In Section~\ref{spatial-entanglement}, we study the spatial structure of entanglement by using the eigenstates associated with the reduced density matrix, as well as a spatially-resolved EE.
We conclude and give an outlook for other topological phases in Section~\ref{conclu}.  

\section{Setup} \label{setup}

\begin{figure}
\centering
  \includegraphics[trim={0 0cm 0 0cm},clip,width=\textwidth*3/5]{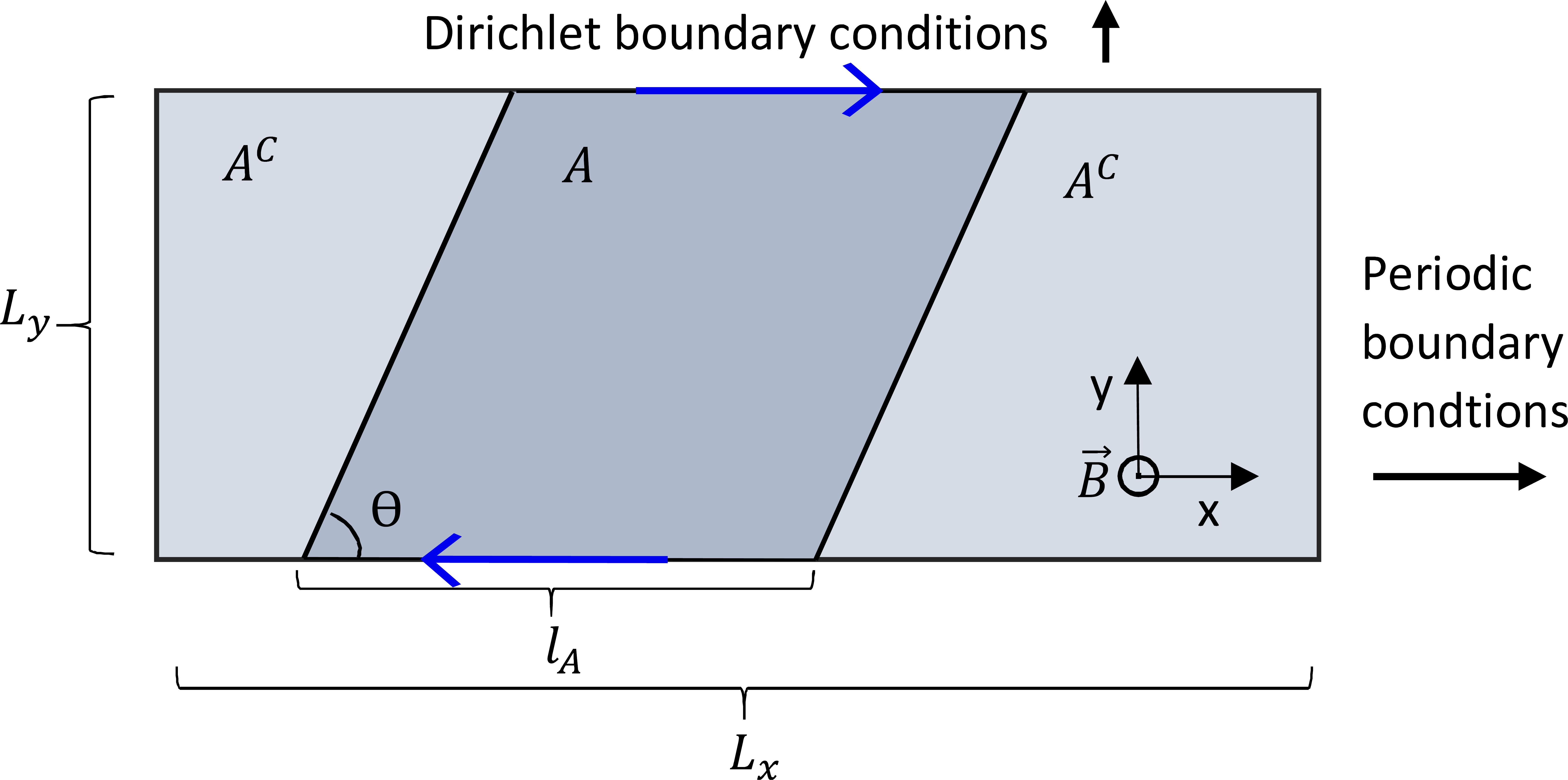}
  \caption{Quantum Hall strip with periodic boundary conditions along $x$, and open or Dirichlet boundary conditions along $y$. The $y=0$ (bottom) and $y=L_y$ boundaries each host a chiral mode, as indicated by the blue arrows. 
  Our results for the reduced density matrix (entanglement entropy, spectrum, and eigenstates) are obtained for subregion $A$; $A^c$ denotes its complement.
  The region has 4 corners intersecting the physical boundaries.}
  \label{illustration_boundary}
\end{figure}

The single-particle Hamiltonian of a quantum Hall strip periodic in $x$ and open in $y$, see Fig.~\ref{illustration_boundary}, is
\begin{equation}
H = \frac{1}{2m_e}(\vec{p}+e\vec{A})^2, \quad y\in [0,L_{y}]
\end{equation}
where $m_e$ is the effective electronic mass of the 2DEG. When $y$ lies outside $[0,L_y]$, we take the potential to be infinite thus imposing Dirichlet boundary conditions $\phi(x,0)=\phi(x,L_y)=0$ on
the electronic wavefunctions. Spin is taken to be fully polarized, and will be omitted from the discussion. Choosing the gauge $\vec{A} = (-By,0)$ preserves translation invariance along $x$ and leads to plane wave solutions in the $x$ direction. We can thus 
decompose the eigenstates of $H$ as $\phi_{n,k}(\vec r)= e^{ikx}f_{n,k}(y)$, where $k$ is a wavevector, and $n$ the Landau level index. The wavevectors are quantized due to the $x$-periodicity, $k=2\pi l/L_x$ with $l\in\mathbb Z$.
The equation for $f_{n,k}(y)$ then reads:
\begin{equation} \label{y-eq}
    \begin{array}{ll}
        \left(-\partial_y^2+( k-y)^2-2m_e E_{n}(k)\right)f_{n,k}(y) = 0, \quad y\in [0,L_{y}]\\
        f_{n,k}(y) = 0, \quad \mbox{otherwise}
    \end{array}
\end{equation}
where we have used the convention $e=\ell_{B} = 1$. In the presence of boundaries, the Landau levels acquire a non-trivial dispersion, $E_n(k)$.
One can now either use the analytical solutions in terms of parabolic cylinder functions 
(also implies numerically solving a transcendental equation, see Appendix~\ref{Solutions_exactes} for more details), or use a numerical approach involving a discrete Fourier expansion. We found it more convenient to opt for the latter, which we now discuss.

We shall work in the LLL, so we set $n=0$. We can expand $f_{0,k}$ using the complete basis of eigenfunctions of the infinite well:
\begin{equation}\label{base_sinus}
f_{0,k}(y)=\sqrt{\frac{2}{L_y}} \sum_{\ell=1}^N b_\ell(k)\sin(\frac{\pi \ell}{L_y } y) 
\end{equation}
where we have truncated the series to the first $N$ terms. We ensure convergence by taking $N$ sufficiently large.
This yields the following discrete eigenvalue problem for the coefficients $b_\ell$:
\begin{align} \label{probleme_valeurs_propres}
\sum_{\ell'} H_{\ell \ell'}(k)b_{\ell'}(k)=2 m E_0(k) b_{\ell}(k) \\
H_{\ell\ell'} =  \frac{2}{L_{y}}\int_0^{L_y}\! dy \sin (\frac{\pi \ell}{L_{y}}y)\;\hat{\mathcal{H}}\;\sin(\frac{\pi\ell'}{L_y}y)
\end{align}
with $\hat{\mathcal{H}} = - \partial_y^2+(k-y)^2$. Plots of the eigenfunctions and energies are presented in Appendix~\ref{Solutions_exactes}. 
Compared to the solutions in terms of parabolic cylinder functions, we found that the decomposition Eq.~\ref{base_sinus}
reduces the computation time and is thus preferable. 
Results concerning convergence of the solutions are presented in Appendix~\ref{Solutions_exactes}.

\section{Entanglement entropy at $\nu=1$} \label{Results}

In order to obtain the reduced density matrix, $\rho_A=\mbox{Tr}_{A^c} \ket{{\rm GS}}\!\bra{{\rm GS}}$, we need to first evaluate the two-point correlator restricted
to region $A$~\cite{Peschel2003,Vidal2003},   
$C^A_{\vec r,\vec r'}=C_{\vec r,\vec r'}$ with $\vec r,\vec r'\in A$: 
\begin{equation}
C_{\vec{r},\vec{r}^\prime} = \bra{{\rm GS}}\psi^\dag(\vec r)\psi(\vec r')\ket{{\rm GS}}, 
\end{equation}
where the electron annihilation operator in the LLL is $\psi(\vec r)= \sum_k \phi_{0,k}(\vec r) c_k$; $c_k$ annihilates 
a fermion with wavevector $k$. $\ket{{\rm GS}}$ is the Slater determinant of electrons with filling $\nu=1$.
One then diagonalizes the two-point correlator:
\begin{align} \label{C-eig}
    \int_{A}d^2\vec{r}^\prime C^A_{\vec{r},\vec{r}^\prime}g_{n}(\vec{r}^\prime) = \lambda_{n} g_{n}(\vec{r})
\end{align}
where $g_n(\vec r)$ is an eigenfunction with corresponding eigenvalue $\lambda_n$. 
The EE is obtained via the standard eigenvalue sum:
\begin{equation}
S_{A} = \sum_{m}h(\lambda_{m}),\;\;\; h(x) = -x\ln{x} - (1-x)\ln{(1-x)}
\end{equation}
The eigenvalues $\lambda_m$ and eigenfunctions $g_n(\vec r)$ encode more information than the EE, and will
be studied in detail below. In dealing with Eq.~(\ref{C-eig}), it is simpler to work in the discrete basis of $k$-modes.
One then needs to find the eigenvalues of the finite-dimensional matrix~\cite{rodriguez2009entanglement,rodriguez2010entanglement}: 
\begin{align} \label{F-matrix}
    F_{lm}=\int_A d^2\vec r\, \phi_{0,k_m}^*(\vec r)\phi_{0,k_l}(\vec r),
\end{align} 
where the number of wavevectors is given by the integer part of $L_x L_y/(2\pi)$. Eq.~(\ref{F-matrix}) is nothing else but the inner product between the $k_l$ and $k_m$ states restricted to subregion $A$. So if $A$ is taken to be the entire system, $F$ becomes the identity matrix and all the eigenvalues become unity, as expected since in that case the EE must vanish.

\subsection{Edge modes}\label{edge_modes}
\begin{figure}
    \hspace*{24mm}{\includegraphics[height=18cm, width=0.75\textwidth, keepaspectratio]{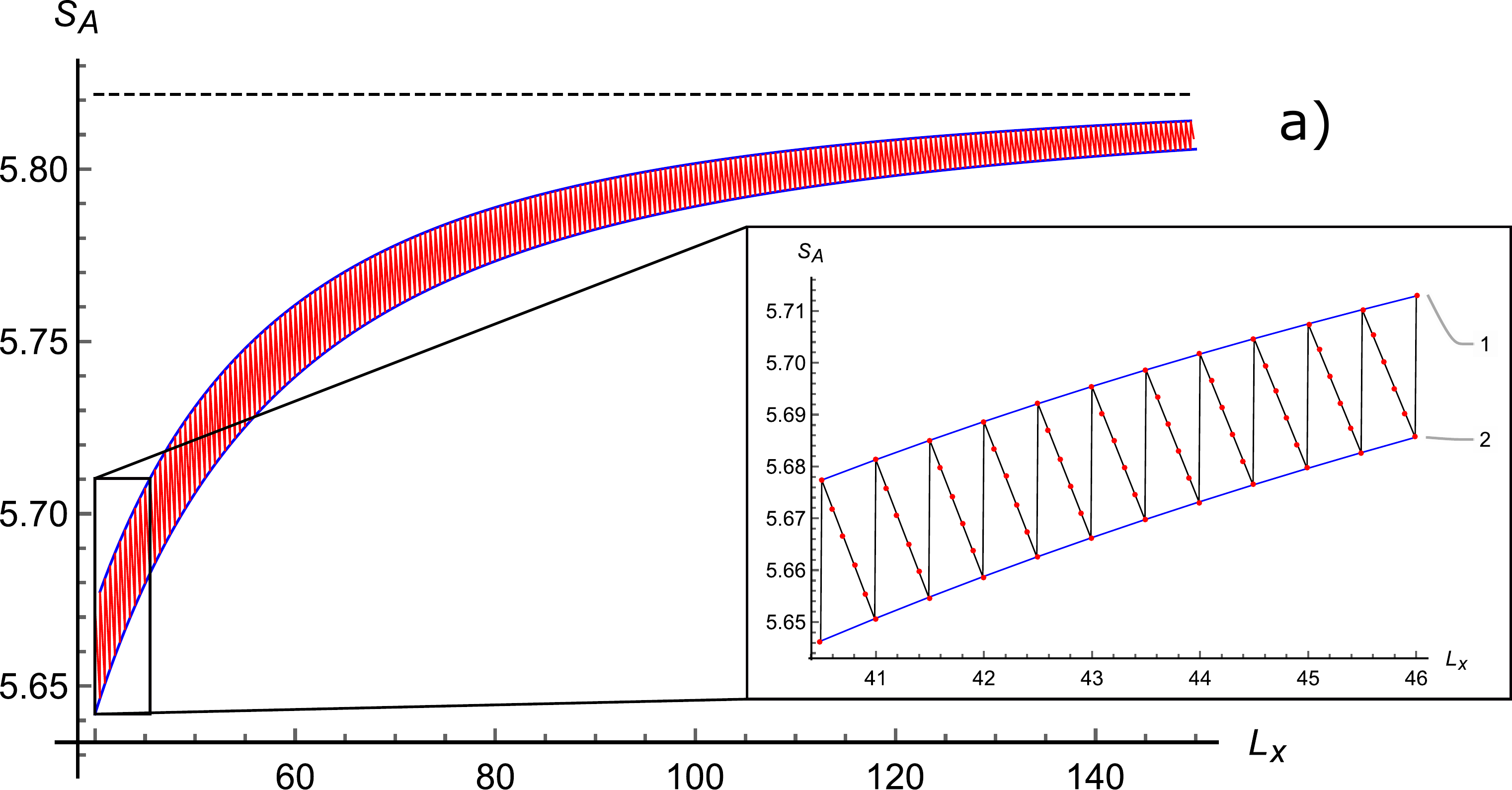}}
    \vspace{2mm}
    \hspace{17mm}{\includegraphics[height=18cm, width=0.6\textwidth, keepaspectratio]{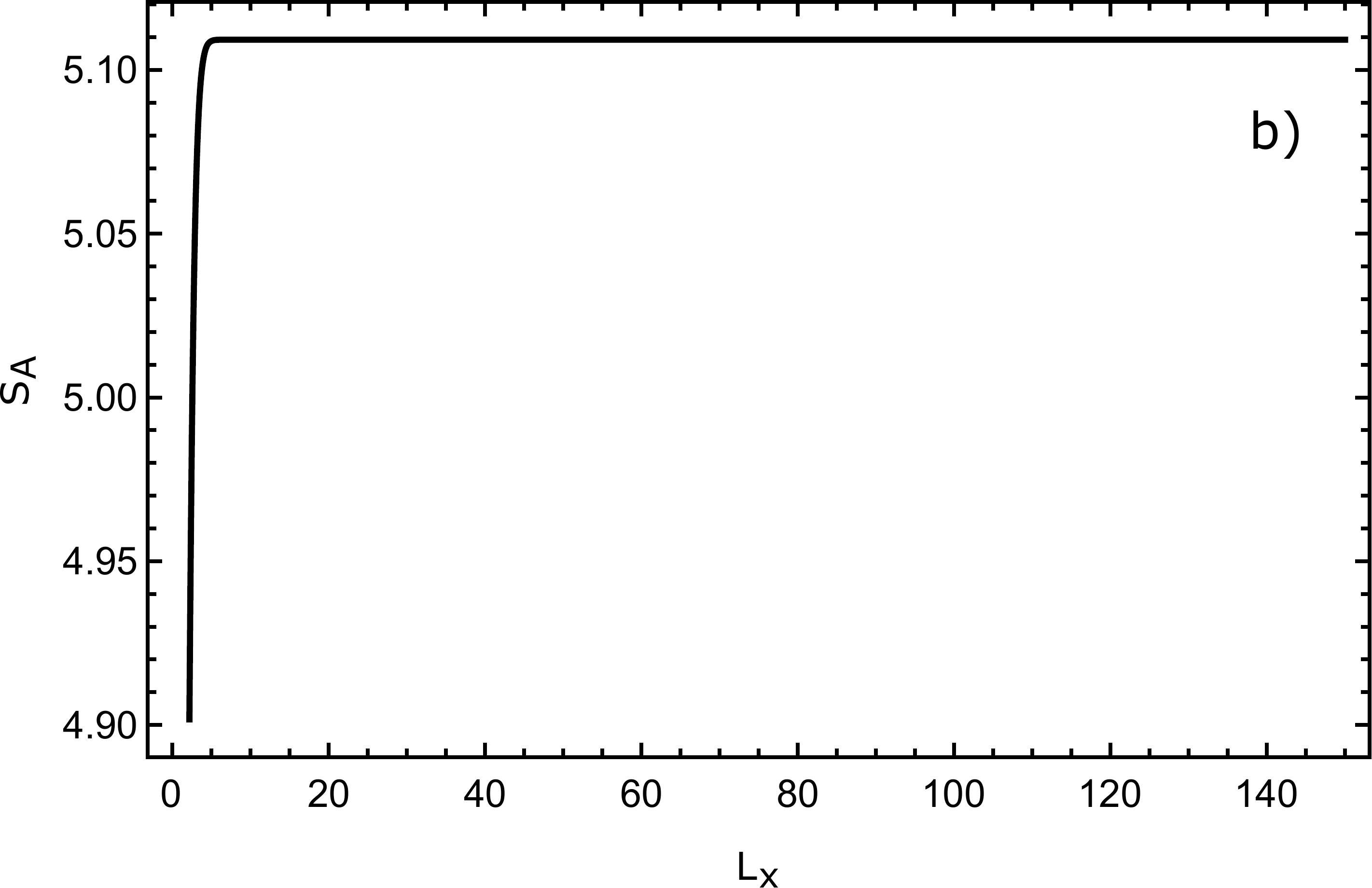}}
    \caption{{\bf a)} Convergence of the entanglement entropy on a strip with open boundary conditions at $y=0,L_y$. The oscillations have a period of $2\pi/L_{y}$. The curve identified by 1 represent the highest possible value of the entanglement entropy in a complete cycle, the curve identified by 2 is the lowest. Both curves converges to the same value 5.822 when $L_{x} \rightarrow \infty $. 
    {\bf b)} Convergence of the entanglement entropy on a torus (periodic boundary conditions in both $x$ and $y$). The following parameters have been used for both panels: $L_{y}$ = $4\pi$, $l_{A}$ = 20, $\theta$ = $\pi/2$}
    \label{fig_compar}
\end{figure}

 The presence of gapless chiral modes on each boundary (see Fig.~\ref{illustration_boundary}) makes the analysis of the EE more involved compared to boundary-less geometries. 
 First, the convergence of the EE requires larger system sizes compared to the torus, as
 can be seen in Fig.~\ref{fig_compar}. In both panels, we keep $L_y$ and $l_A$ fixed for $\theta=\pi/2$, and increase $L_x$. The
 low energy modes arising from the boundary lead to much slower convergence to the asymptotic $L_x=\infty$ EE, represented by
 the dashed line in the top panel of Fig.~\ref{fig_compar}. In the bottom panel, we show the convergence for the torus geometry with a region $A$ of the same size, and there the EE converges to its asymptotic value after only a few magnetic lengths. 
 As a second new feature compared to the torus, we observe oscillations of the EE when increasing $L_{x}$, clearly shown
 in the inset of the top panel of Fig.~\ref{fig_compar}. The oscillations have period $2\pi/L_y$ and can be understood as follows.
 One has to fill $\lfloor{L_{x}L_{y}/(2\pi)\rfloor}$ modes to obtain the $\nu=1$ groundstate. Consider now that $L_x L_y/(2\pi)$
 is an integer. If we then progressively increase $L_x$ (keeping $L_y$ fixed), the number of filled $k$-modes remains constant. However, the spacing between the modes, $2\pi/L_x$, decreases. Due to the confining potential, this means that the occupied modes move away from the boundaries to lower the system's energy. The edge modes become slightly more ``bulk-like'', and are thus less effective at contributing to the entanglement between $A$ and its complement. This explains the progressive reduction of the EE within one oscillation period. Once $L_x L_y/(2\pi)$ reaches the next integer, as $L_x$ increases by $2\pi/L_y$, a new mode maximally close to the boundary is added and the EE jumps back up. The cycle repeats itself with a reduced oscillation amplitude as $L_x$ grows larger since such finite size effects become less important as the thermodynamic limit is approached. 
 
\begin{figure}
\centering
  \includegraphics[width=\textwidth*2/3]{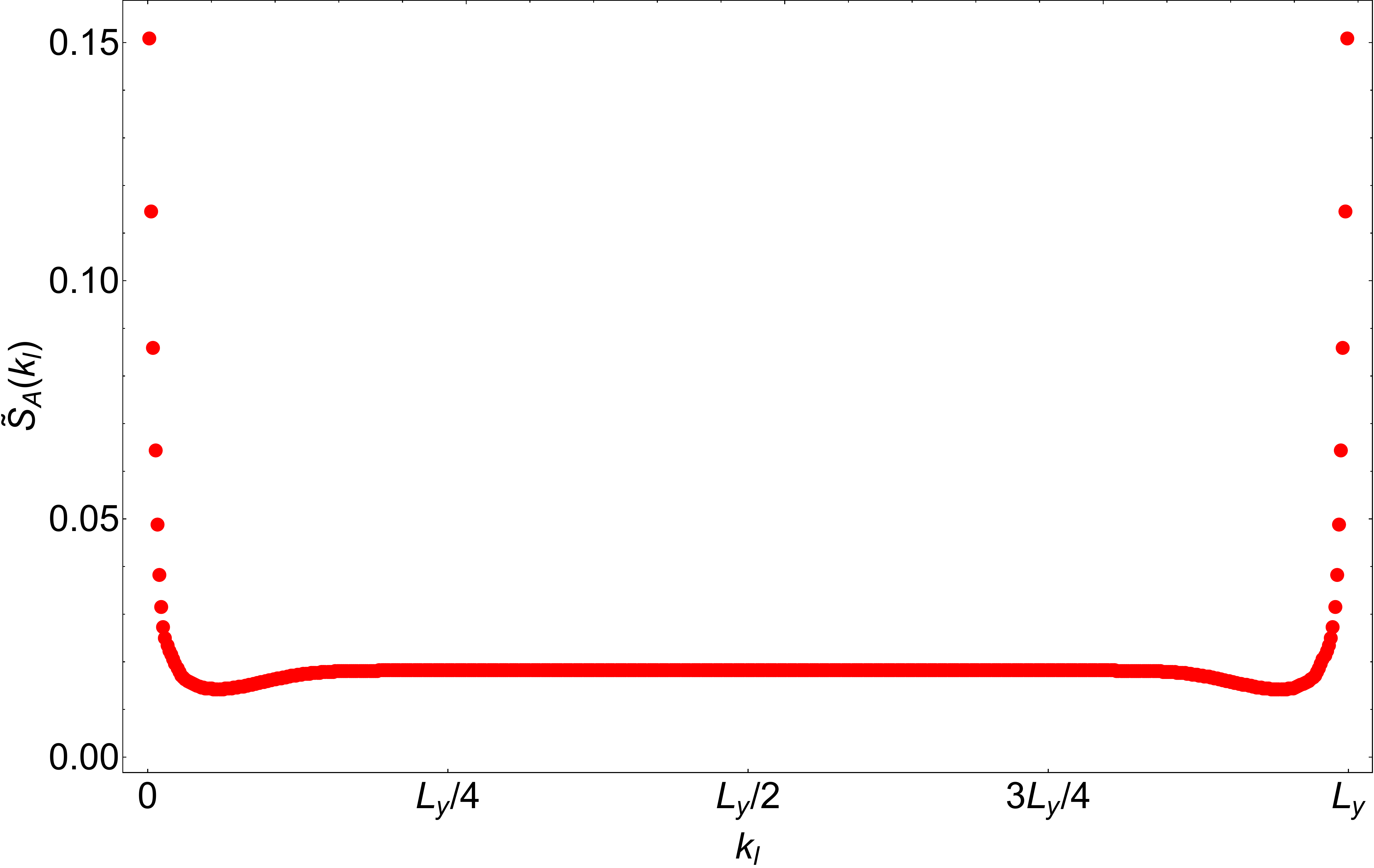}
  \caption{Entanglement entropy per mode $\tilde S_A(k_l)$
  for $\theta = \pi/4$, $L_x = 200$, $l_A$ = 20, $L_y$ = $6\pi$. $\tilde S_A(k_l)$ becomes maximal for modes located near the boundaries.}
  \label{entropie_par_mode}
\end{figure}

In order to provide a more quantitative perspective on the above discussion,
we can determine the contribution of the mode with wavevector $k_l$ to the EE, which we call $\tilde S_A(k_l)$: 
 \begin{align}
 \tilde{S}_A(k_l) = \sum_{m} h(\lambda_{m}) |V_{m,l}|^2 \\
 \label{entropi_mode} 
 S_A = \sum_{k_{l}}\tilde{S}_A(k_l)
\end{align}
where $\vec{V}_{m}$ is the normalized eigenvector associated with the eigenvalue $\lambda_m$ of the correlator $C^A_{\vec r,\vec r'}$: $g_m(\vec r)=\sum_{k_l} V_{m,l}\phi_{0,k_l}(\vec r)$. 
If one sums the entropy-per-mode over all the modes of the LLL, one recovers the full EE, Eq.~(\ref{entropi_mode}). In Fig.~\ref{entropie_par_mode}, we clearly see that $\tilde S_A(k_l)$ is maximal for modes located near the boundaries of the strip. Such propagating modes naturally play an important role in mediating entanglement between $A$ and its complement. In contrast, 
for a mode deep in the bulk, $\tilde S_A(k_l)$ is 
much smaller as expected. However, in the thermodynamic limit the number of such bulk modes diverges (see the wide plateau in Fig.~\ref{entropie_par_mode}), and they dominate the EE through the area law contribution.

\subsection{Area law and subleading logarithm}
Using the methods described in Section~\ref{Results} and taking the limit $L_{x} \rightarrow \infty$ (see Fig. \ref{extrapolation_Lx} for more details), one can extract the
area law  by keeping $l_{A}$ and $\theta$ constant while changing $L_{y}$:
\begin{align}
    S_A=\coeffD\, \frac{P_A}{\ell_B} +\cdots,
\end{align}
where the perimeter of the region is $P_A=2L_y/\sin\theta$, and the ellipsis denote terms subleading as $\ell_B\to 0$. 
Note that we have reinstated $\ell_B$. 
We numerically determined the area law coefficient to be $\coeffD$ = $0.203291\pm 10^{-6}$, which agrees up to 6 digits with previous computations for the $\nu=1$ state on a geometry without physical boundaries, such as the torus \cite{rodriguez2009entanglement}. In the thermodynamic limit,
$\beta$ is given by an integral involving the error function~\cite{rodriguez2009entanglement,rodriguez2010entanglement}, and can be evaluated with arbitrary precision: $\beta^{\rm thermo}\simeq 0.20329081323$.

\begin{figure}
\centering
  \includegraphics[width=\textwidth*2/3]{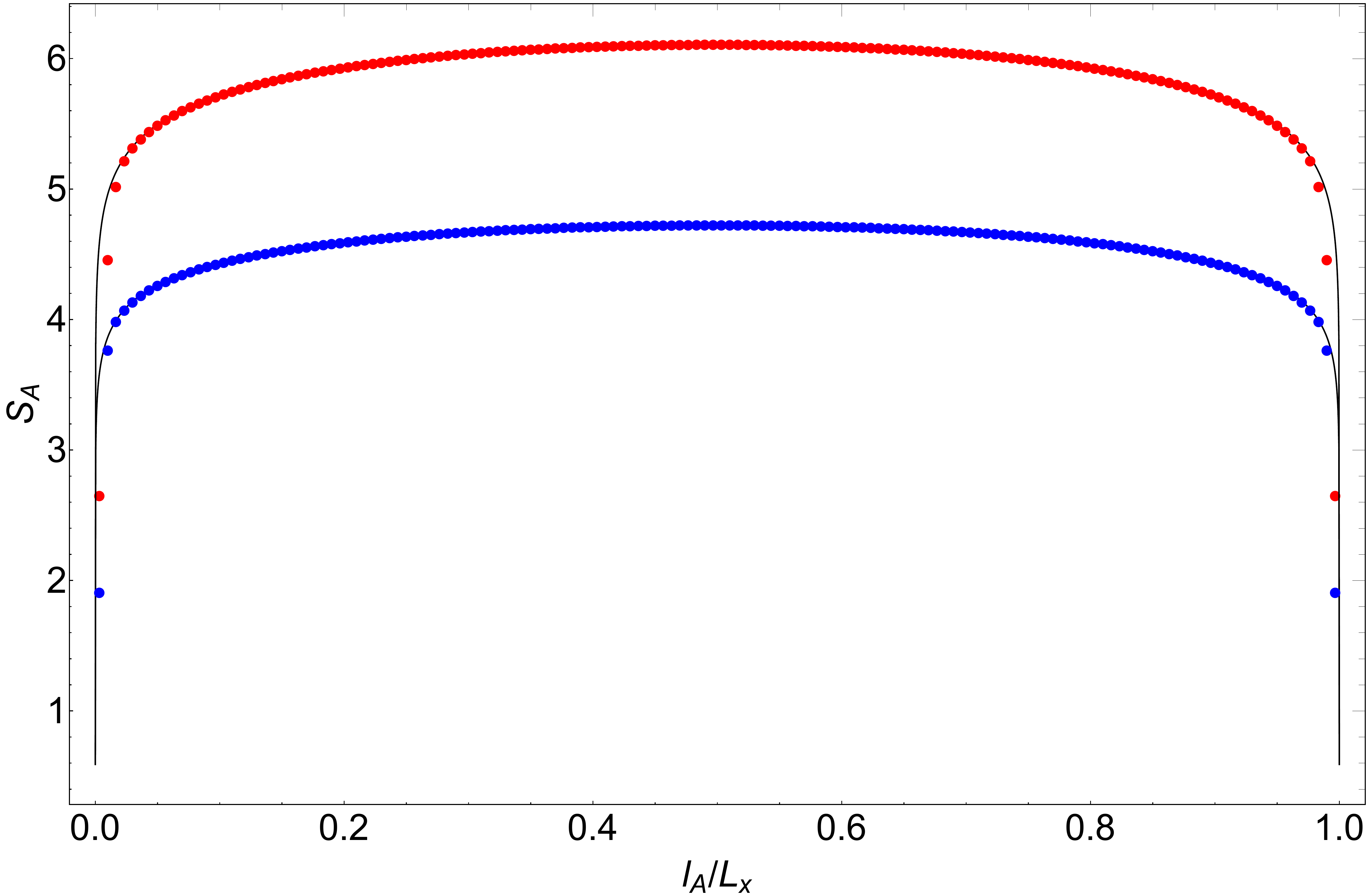}
  \caption{Numerical results for the entanglement entropy $S_{A}$ (red dots) and the 2nd R\'enyi entropy $S_{A,2}$ (blue dots) as a function of $l_A/L_x$ for $\theta = \pi/2$, 
  $L_{x} = 150$, $L_y = 6\pi$. The solid black lines correspond to the 1-dimensional conformal field predictions with central charge $c=1$.}
  \label{divergence_fit1D}
\end{figure}

By computing the EE for different values of $l_{A}$ while keeping $L_{y}$ and $\theta$ fixed, we obtain the following dependence, as shown in Fig.~\ref{divergence_fit1D}: 
\begin{equation} \label{CFT}
S_{A} =\beta \frac{P_A}{\ell_B}+ 2\times \frac{c}{6}\ln{\left(\frac{L_{x}}{\pi \ell_{B}}\sin{\left(\frac{\pi l_{A}}{L_{x}}\right)}\right)} + C'
\end{equation}
where we have again reinstated the magnetic length $\ell_B$. 
In Fig.~\ref{divergence_fit1D}, the black curve corresponds to $c=1$.
The logarithmic term, which is subleading as we take $\ell_B\to 0$, holds when $l_A \gg \ell_{B}$ and does not depend on $\theta$ or $L_y$. It correspond to the celebrated expression for the EE of a segment in a non-chiral CFT defined on a circle of length $L_x$,
with periodic boundary conditions \cite{CC2004}.
The central charge that we find, $c=1$, corresponds to a free Dirac fermion CFT as expected from the boundary modes of the $\nu=1$ integer quantum Hall state. One can also bosonize the fermionic CFT to get a non-chiral Luttinger liquid. 
The edge-mode contribution to the EE can be decomposed into a contribution coming from each boundary, which hosts a \emph{chiral} mode.
This explains why we explicitly wrote a factor of 2 in front of the logarithm in Eq.~(\ref{CFT}).
Such a subleading logarithmic term in 
EE in two spatial dimensions was previously observed in the numerical study of FQH states with various interfaces~\cite{Varjas2013,crepel_2019,crepel_2019_2,crepel_microscopic_2019}. Interestingly, it was also obtained in gapless systems with a zero mode, such
as free boson or Dirac fermion CFTs in two spatial dimensions with periodic boundary conditions \cite{Chen2017}. 

\subsection{Corner entanglement}
We can now examine the contributions that arise from corners in the entangling region. Indeed, we see
from Fig.~\ref{illustration_boundary} that the entangling surface $\partial A$ intersects the physical boundary four times, yielding two corners with angle $\theta$, and two others with complementary angle $\pi-\theta$. Such boundary corners
will contribute a constant to the EE ($C'$ in Eq.~(\ref{CFT})), analogously to what happens with corners in the bulk:
\begin{align}
    S_{A} =\beta \frac{P_A}{\ell_B}+ \frac{c}{3}\ln{\left(\frac{L_{x}}{\pi \ell_{B}}\sin{\left(\frac{\pi l_{A}}{L_{x}}\right)}\right)} - \sum_{i=1}^4 b(\theta_i) + \cdots
\end{align}
where the sum is over all the boundary corners of $A$. The terms denoted by the ellipsis include contributions that vanish as $\ell_B\to 0$, as well as a non-vanishing term that is angle-independent. We define our $b(\theta)$ such that $b(\pi/2)=0$, i.e.\ perpendicular corners do not contribute to $b(\theta)$. This ambiguity does not occur for bulk corners.
Indeed, for a region that is far from physical boundaries (if any), the EE in the presence of a bulk corner is $S_A= \beta P_A/\ell_B - a(\theta)$, where
$\theta$ is the corner's opening angle. For integer quantum Hall states, one immediately finds that $a(\pi)=0$ vanishes when the corner
disappears~\cite{rodriguez2010entanglement}.

\begin{figure}
\centering
  \begin{subfigure}[b]{0.52\textwidth}
          \includegraphics[width=\textwidth]{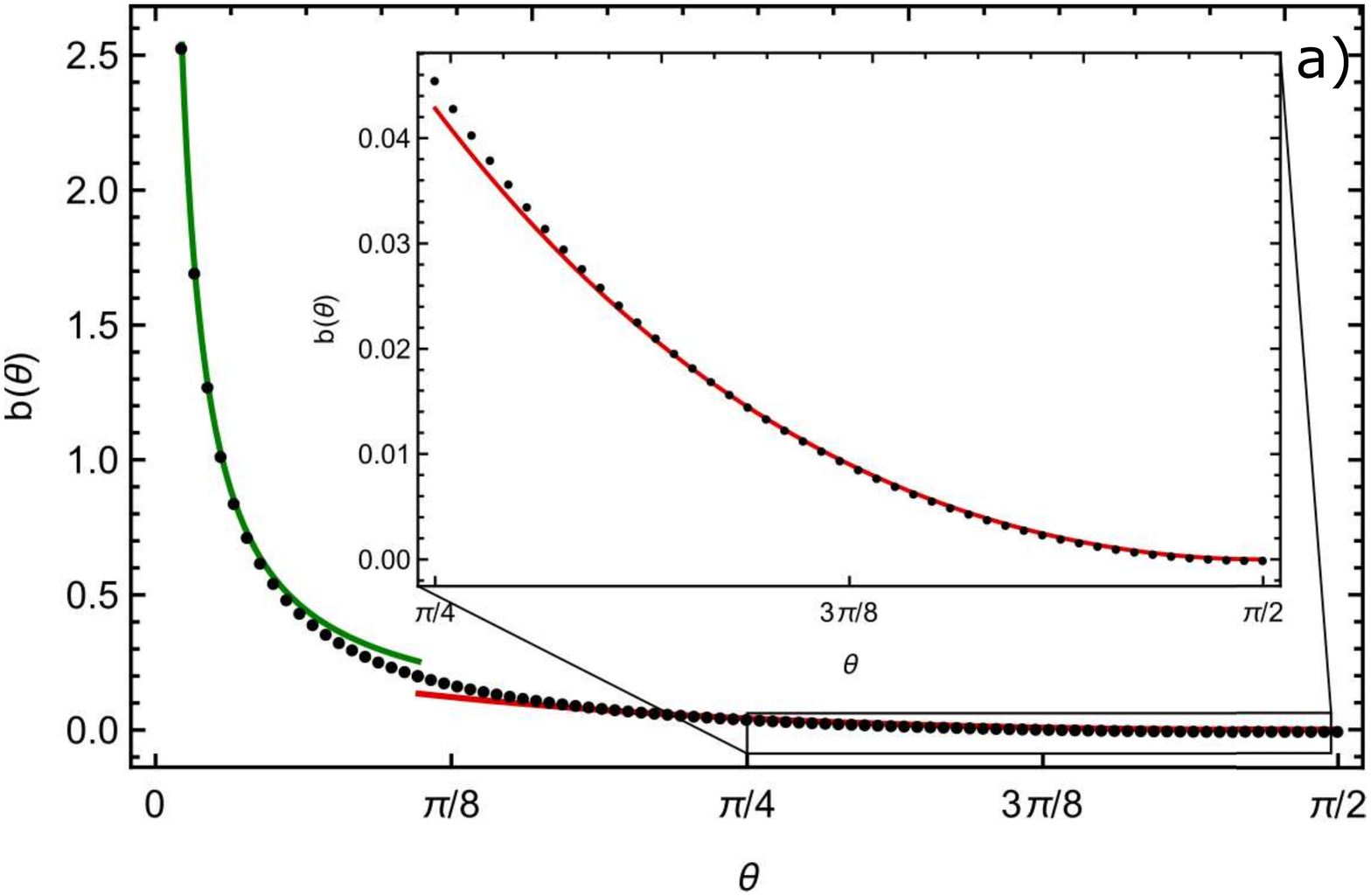}
    \label{fonction_sinus}
  \end{subfigure}
  \begin{subfigure}[b]{0.47\textwidth}
      \includegraphics[width=\textwidth]{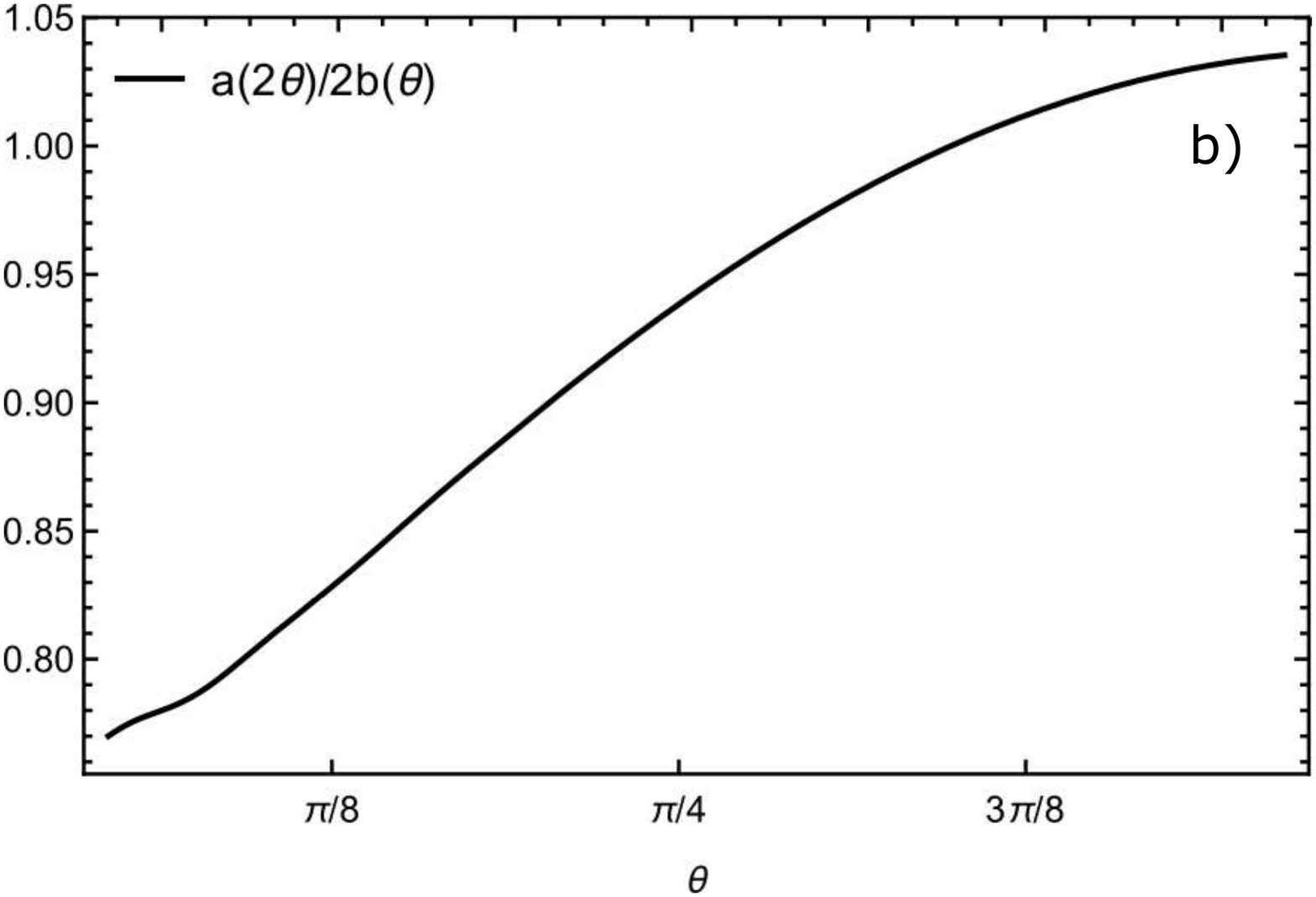}
    \label{fig:2}
  \end{subfigure} 
  \caption{$\bf{a)}$ Boundary corner function $b(\theta)$ for angles ranging from $2^\circ$ to $90^\circ$. The red curve shows the fit for angles near $\pi/2$, Eq.~(\ref{exp-pi2}), and the green curve shows the $1/\theta$ fit at small angles. $\bf{b)}$ Comparison of $2b(\theta)$ to the bulk corner function $a(2\theta)$. Calculations performed with the following parameters: $L_{x}$ $\rightarrow$ $\infty$, $l_{A}  = 20$ and the two values of $L_{y}$ being respectively $4\pi$ and $6\pi$.}
  \label{btheta}
\end{figure}

The boundary corner function satisfies $b(\theta)=b(\pi-\theta)$ since for a pure state $S_A=S_{A^c}$. As such, we only need to
study $b(\theta)$ for $0< \theta\leq \pi/2$. For our parallelogram geometry (Fig.~\ref{illustration_boundary}), we thus have
\begin{align}
    \sum_{i=1}^4 b(\theta_i) = 4 b(\theta)
\end{align}
By varying $\theta$ we can numerically determine the boundary corner function. The result is shown in the left panel of
Fig.~\ref{btheta}. We see that it is positive and monotonically decreasing on $(0,\pi/2]$. In addition, it has a divergence as the angle vanishes, and goes to zero in the opposite limit, $b(\pi/2)=0$. 

We now analyze the behavior of $b(\theta)$ more closely. First, we expect $b(\theta)$ to
be analytic near $\pi/2$ since nothing singular happens with region $A$ as $\pi/2$ is crossed. In addition, the reflection symmetry of $b(\theta)$ about $\pi/2$ can be used to conclude that only even powers of $(\theta-\pi/2)$ appear:
\begin{equation}\label{exp-pi2}
b(\theta) = \sigma\cdot\left(\theta-\frac{\pi}{2}\right)^{2} + \tilde{\sigma}\cdot \left(\theta-\frac{\pi}{2}\right)^{4}
+ O\!\left(\left(\theta-\frac{\pi}{2}\right)^6\right)
\end{equation}
The red line in the left panel of Fig.\ref{btheta} is a fit to the above expression,
which shows that the boundary corner function indeed obeys Eq.~(\ref{exp-pi2}) near $\pi/2$; the inset shows the region near $\pi/2$ in more detail. The corresponding coefficients $\sigma$ and $\tilde\sigma$ are
shown in Table~\ref{afit}.

In the opposite limit of small angles, we expect the $1/\theta$ divergence seen in other systems such as CFTs in two spatial dimensions~\cite{Seminara2017,Berthiere2019-1,Berthiere2019}
\begin{equation}
b(\theta\to 0) = \frac{\kappa}{\theta}
\end{equation}
The corresponding fit is shown in green in the left panel of Fig.~\ref{btheta}. It perfectly matches the data as $\theta\to 0$.
The sharp-limit coefficient $\kappa$ is given in Table~\ref{afit}. 

It was recently found \cite{Berthiere2019} that the boundary corner function $b(\theta)$ is related to the bulk corner function $a(\theta)$ in various 2d quantum systems. More precisely, for certain quantum field theories (not necessarily relativistic) and boundary conditions, it was found that~\cite{Berthiere2019} 
\begin{align} \label{bulk-bdy}
	2b(\theta)=a(2\theta). 
\end{align}
Examples include scalar and Dirac fermion CFTs, certain CFTs with holographic duals~\cite{Seminara2017}, and a gapless Lifshitz theory with dynamical exponent $z\!=\!2$.
Since $a(\theta)$ is known numerically for the $\nu=1$ groundstate \cite{rodriguez2010entanglement}, we can verify if Eq.~(\ref{bulk-bdy}) holds. The right panel of Fig.~\ref{btheta} shows the ratio $a(2\theta)/(2b(\theta))$. At large angles it is near $1.04$, while it goes to $0.76$ at small angles. We thus see that the simple relation between $b(\theta)$ and $a(\theta)$ is not obeyed for the $\nu=1$ integer quantum Hall state with Dirichlet boundary condition. However, the ratio is near unity for a wide range of angles, and equals 1 for a unique angle near $3\pi/8$. It would be of interest to understand why the previously observed bulk-boundary correspondence for the corner term is almost obeyed. One possibility is that a more general form of the bulk-boundary relation is obeyed~\cite{Berthiere2019}: $a(2\theta)=b^{\mathcal B}(\theta)+b^{\mathcal B'}(\theta)$, where $\mathcal B$ and $\mathcal B'$ correspond to \emph{different} boundary conditions. If $\mathcal B$ is taken to be Dirichlet, it remains to be seen if there exist a partner boundary condition $\mathcal B'$ so that the generalized bulk-boundary relation is obeyed. 
\begin{table}
    \centering
    \begin{tabular}{l||ll}
      ${}$ & Estimate &  Error \\
    \hline
     $\kappa$ & $0.0886$ & $0.0004$  \\
     $\sigma$ & $0.0543$ & $0.0002$  \\
     $\tilde{\sigma}$ & $0.0278$ & $0.0006$  
    \end{tabular}
    \smallskip
    \caption{Numerical values and uncertainties of the fitting parameters in the $\theta\to0$ and $\theta\to\pi/2$ limits
    of $b(\theta)$.} 
    \label{afit}
\end{table}

\subsubsection{R\'enyi entropies}
\begin{figure}
\centering
  \includegraphics[width=\textwidth*2/3]{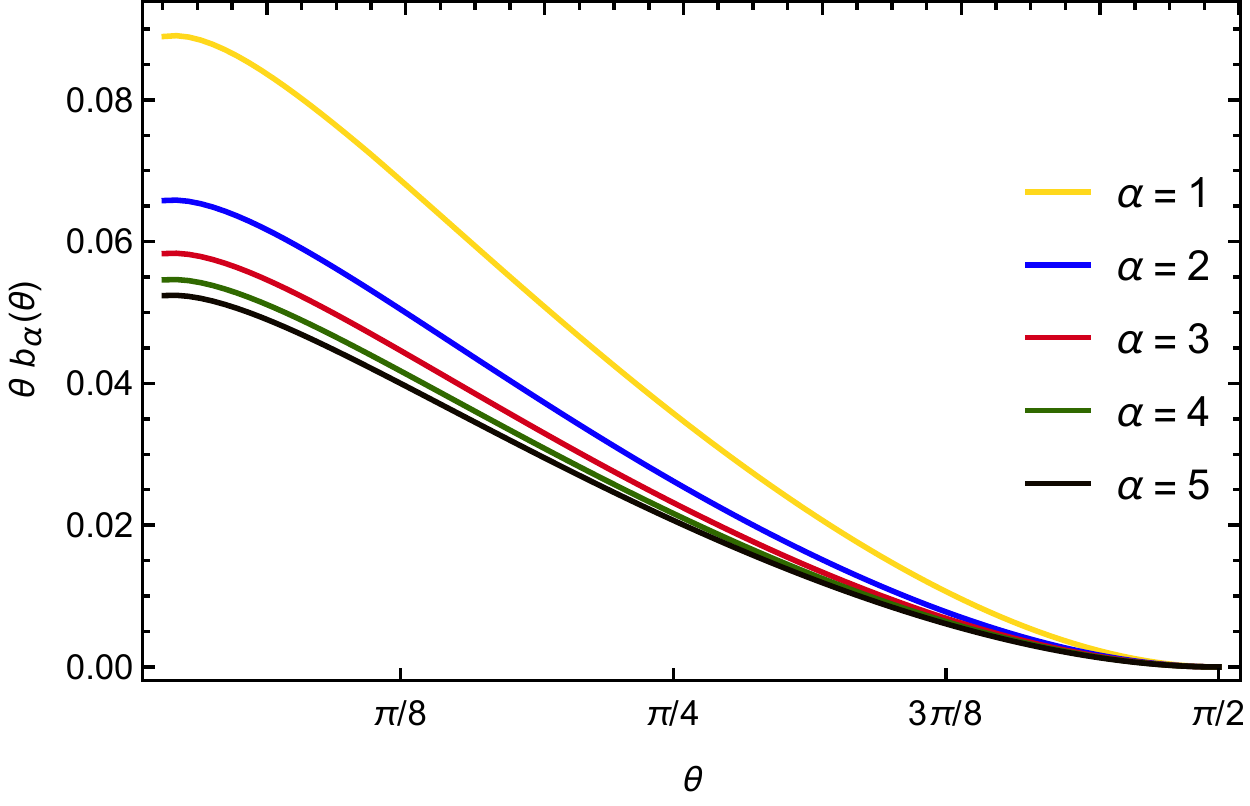} 
  \caption{R\'enyi corner function, $\theta b_{\alpha}(\theta)$, for angles ranging from $2^\circ$ to $90^\circ$ computed for different values of the R\'enyi index $\alpha$. Calculations are performed in the limit $L_{x}$ $\rightarrow$ $\infty$ with $l_{A}  = 20$ and the two values of $L_{y}$ being respectively $4\pi$ and $6\pi$.}
  \label{courbe_a(alpha)}
\end{figure}

We now study the R\'enyi entanglement entropies:
\begin{equation}S_{A,\alpha} = \frac{1}{1-\alpha}\ln{\Tr(\rho^{\alpha}_{A})},
\end{equation}  
which reduce to the von Neumann EE as the R\'enyi index approaches unity, $\alpha\to 1$. 
We find that the prefactor of the edge-mode contribution $\tfrac{c}{3}$ in Eq.~(\ref{CFT}) is replaced by $\tfrac{c}{6}(1+\alpha)$, which is the expected result for a CFT in 1 spatial dimension~\cite{calabrese2004entanglement}. Fig.~\ref{divergence_fit1D} shows the corresponding fit for $\alpha=2$.
Next, let us investigate the R\'enyi dependence of the corner function $b_\alpha(\theta)$.
Fig.~\ref{courbe_a(alpha)} shows $\theta\, b_\alpha(\theta)$ for different values of the R\'enyi index. 
This way of plotting makes the $1/\theta$ divergence of $b$ manifest. We observe that $b_\alpha(\theta)$ decreases monotonically with increasing $\alpha$. This was also observed to be the case for CFTs in 2 spatial dimensions with Dirichlet or mixed boundary conditions~\cite{Berthiere2019}. For the free scalar CFT in 2 spatial dimensions with Neumann boundary conditions, it is $|b_\alpha(\theta)|$ that decreases with increasing $\alpha$.
We further find that the dependence on the index $\alpha$ cannot be separated from the angle dependence, just as for many gapless 2d systems with~\cite{Berthiere2019}
and without~\cite{Bueno2015} boundaries. Indeed, we find that the ratio $b_1/b_2$ takes the following values at selected angles:  
\begin{align}
    \frac{b_1(45^\circ )}{b_2(45^\circ)}=1.365\,,\quad
     \frac{b_1(60^\circ)}{b_2(60^\circ)}=1.369\,,\quad
      \frac{b_1(80^\circ)}{b_2(80^\circ)}=1.372\,,
\end{align}
We see that the ratio is not constant as a function of $\theta$, but it does not change much. Interestingly, the ratio is not too far from what we would get if $b_\alpha(\theta)=(1+\tfrac{1}{\alpha})f(\theta)$, which is a guess based on the R\'enyi dependence for CFTs in 1 spatial dimension.
This naive guess gives a ratio equal to $4/3$ independently of the angle, not too far off from what we observe. Similar behavior is obtained for the free scalar CFT with Dirichlet boundary condition in 2 spatial dimensions~\cite{Berthiere2019-1,Berthiere2019}, where the ratio is near $4/3$ but slightly above, and also varies with the angle. 

\subsection{Entanglement spectrum}
We now examine the eigenvalues $\lambda_n$ of the two-point spatial correlator $C_{\vec r,\vec r'}^A$, which are directly related to the single-particle entanglement spectrum through $\lambda=n_f(\epsilon)$, where $\epsilon$ is the corresponding pseudo-energy. We note that since the entangling region breaks spatial translations in both directions, we cannot use momentum to label the eigenvalues. 

\begin{figure}
\centering
  \includegraphics[width=\textwidth*2/3]{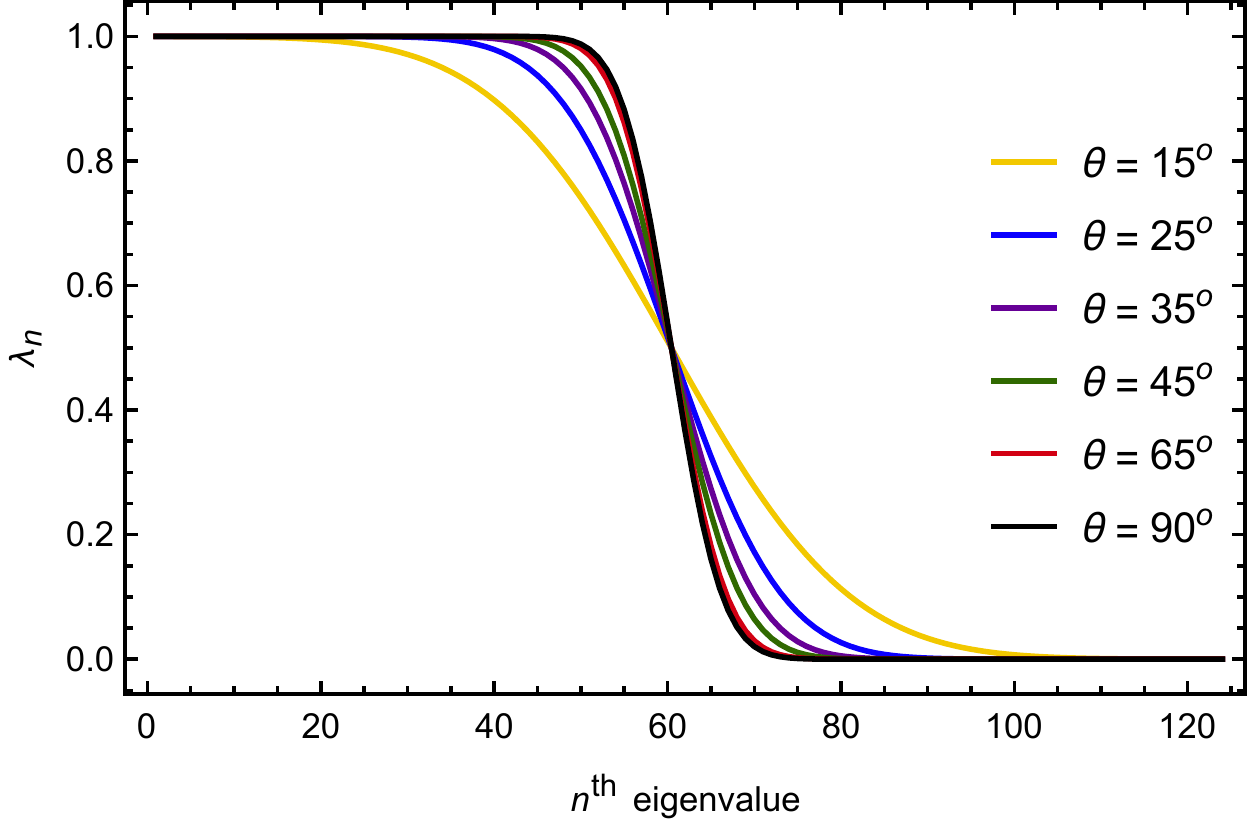}
  \caption{Spectrum of the two-points correlator for different values of $\theta$ with $L_{x} = 150$, $L_{y}$ = $4\pi$, $l_{A}$ = 20.}
  \label{ES} 
\end{figure}

The numerically determined spectrum is shown in Fig.~\ref{ES} for different values of $\theta$. The dominant eigenvalues are around $\lambda=1/2$ since they contribute the most to the EE. We see that the spectrum gets progressively flatter in the middle region as the angle decreases, meaning that more eigenvalues become important for the reduced density matrix.  
The other parameters modify the spectrum in a relatively predictable way. Increasing the value of $L_{y}$ increases the density of points along the spectrum. Increasing $L_{x}$ leads to more points in the region where $\lambda\simeq 0$ without changing the density of points. Finally, changing $l_{A}$ changes the position of the sloping part.
For example, taking $l_{A}$ to be $L_{x}/2$ leads to a symmetric spectrum. 

\section{Spatial distribution of entanglement} \label{spatial-entanglement}

\begin{figure}
\centering
  \includegraphics[width=\textwidth*3/4]{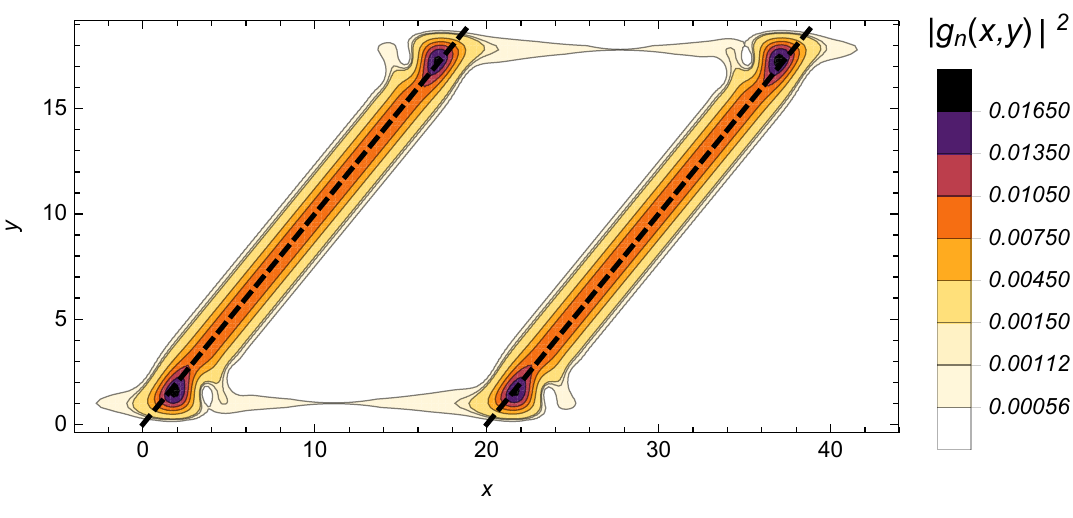}
  \caption{Contour-plot of the eigenvector obtained by diagonalization of the two point correlator for $\lambda_{m}$ $\simeq{}$ 0.5. The black dashed lines are the boundary between $A$ and $A^c$. $ L_y = 6\pi$, $ l_A = 20$, $L_x=200$, $\theta=45^\circ$.}
  \label{fonction_propre}
\end{figure}

We now turn to the eigenfunctions of the two point correlator, $g_n(\vec r)$, which are closely related to the entanglement Hamiltonian. In fact, they can help us ``see'' where does the entanglement between $A$ and $A^c$ come from. 
Let us focus on the eigenfunction corresponding to the eigenvalue that contributes the most to the EE, $\lambda_{m} = 1/2$, or the one nearest to $1/2$.
In Fig.~\ref{fonction_propre}, we show its norm squared, $|g_m(\vec r)|^2$.
It has largest norm along the entangling surface separating $A$ from its complement. In addition, it has global maxima located near the 4 corners.
Further, we see a finite weight along the physical boundaries, arising from the chiral edge modes. 

\begin{figure}
\centering
  \includegraphics[width=\textwidth*3/4]{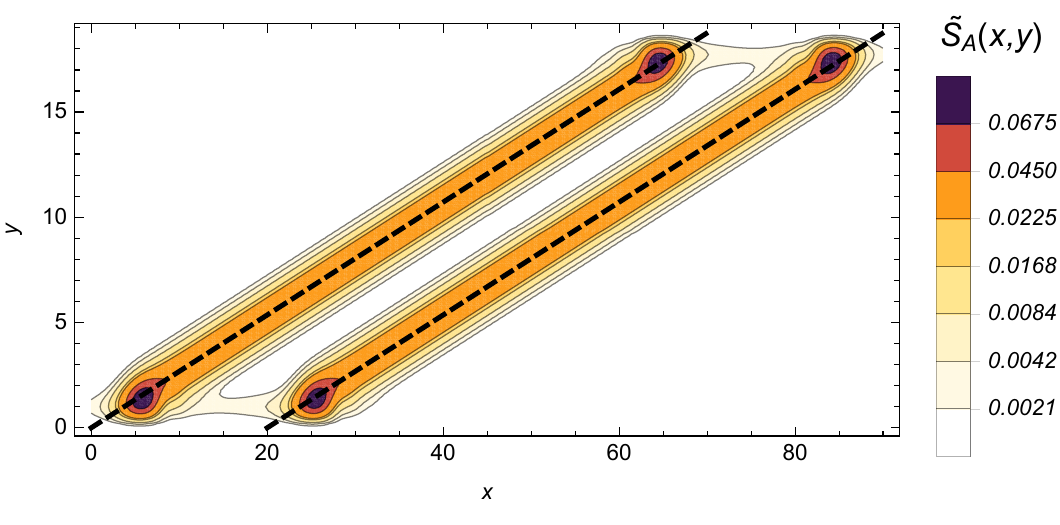}
  \caption{Spatial distribution of the entanglement entropy, $\tilde{S}_A(\vec r)$. The black dashed lines are aligned with the entangling surface, $\partial A$.
    The following parameters have been used: $L_y =6\pi$, $l_A=20$, $L_x=200$, $\theta=15^{\circ}$.} 
  \label{Sxy}
\end{figure}

We can use the eigenfunctions to define the following quantity (see Refs.~\cite{Botero2004,Nozaki2013,ChenVidal2014}):  
\begin{equation}
\tilde{S}_A(\vec r) = \sum_{n} h(\lambda_{n})\abs{g_{n}(\vec r)}^2 
\end{equation}
which, when integrated, gives the total EE, $S_A$. As such, $\tilde S_A(\vec r)$ measures how the EE is spatially distributed. 
In Fig.~\ref{Sxy}, we show $\tilde S_A(\vec r)$ for a particular geometry. Not surprisingly we see that it follows the dependence of $|g_n|^2$ shown in Fig.~\ref{fonction_propre}. We clearly see that the EE is dominated the contributions across the boundary between $A$ and its complement. The contributions from the chiral edge modes and the corners can also be seen. 

\section{Conclusions \& Outlook} \label{conclu}
We have studied the groundstate of the IQH system at $\nu=1$ in the presence of hard-wall boundaries through the reduced density matrix of a spatial region. The region that was studied is a parallelogram that intersects the physical boundary, leading to 4 corners. We found that the EE contains a logarithmic contribution coming from the chiral edge modes, and matches the 1d CFT prediction, Eq.~(\ref{CFT}). 
We uncovered an additional contribution to the EE that arises due to to the 4 boundary corners, $-4b(\theta)$. 
We characterized the angle-dependence of $b(\theta)$, and compared it to the bulk corner EE. We found that the relation Eq.~(\ref{bulk-bdy}) observed in
numerous quantum critical states, is almost obeyed, but deviations do appear. We further analyzed the spatial structure of entanglement via the eigenstates associated with the reduced density matrix, and constructed a spatially-resolved EE. The influence of the physical boundary and the region's geometry on the reduced density matrix was clarified, as shown in Figs.~\ref{fonction_propre}-\ref{Sxy}. 

It would be of interest to study these geometrical aspects of entanglement in FQH states, and in more general topological phases.
For one, we expect the logarithmic term in the EE arising from the edge modes to appear whenever protected gapless modes are present. As a simple example,
it would be present in a 2d quantum spin Hall insulator, which has protected counter-propagating edge modes on each boundary.
The subleading logarithmic term was previously observed in the numerical study of FQH states with various interfaces~\cite{Varjas2013,crepel_2019,crepel_2019_2,crepel_microscopic_2019}.
In addition, it would be of interest to understand what physical information is encoded in angle-dependence of $b(\theta)$, and analyze it for other topological states.
In CFTs in 2 spatial dimensions, it was shown that the function $b(\theta)$ contains important information such as certain conformal anomaly coefficients as well as stress tensor central charges~\cite{Fursaev2016,Seminara2017,Berthiere2019}.  

\begin{acknowledgments}
We would like to thank C.~Berthiere, L.~Fournier, G.~Sierra, B.~Sirois, and J.-M.~St\'ephan for useful discussions. 
P.-A.B.\ would like to thank Universit\'e de Montr\'eal for warm hospitality during the completion of this project.
This work was funded by the Fondation Courtois, a Discovery Grant from NSERC, a Canada Research Chair, and a 
``\'Etablissement de nouveaux chercheurs et de nouvelles chercheuses universitaires'' grant from FRQNT.\\
\emph{Note added}---Shortly before completion of this paper, we became aware of a related work by J.-M.~St\'ephan and B.~Estienne, which partially overlaps with our results.
\end{acknowledgments}

\appendix
\section{Entanglement entropy per mode}  \label{Justification_figure}
The entanglement entropy for our IQH state takes the following form:
\begin{align} \label{ee2}
  S_A = \mbox{Tr}\left( h(P^{-1}F P)\right)
\end{align}
where $F_{lm}$ is the two-point correlator expressed in mode space, Eq.~(\ref{F-matrix}), and $P$ is the matrix that diagonalizes $F$.
By cyclicity of the trace, Eq.~(\ref{ee2}) is equivalent to 
\begin{align}
  S_A = \mbox{Tr}\left(Ph(P^{-1}F P)P^{-1}\right)
\end{align}
The matrix inside the trace takes the following form:  
\begin{align} \label{mat1}
  Ph(P^{-1}FP)P^{-1} = \begin{bmatrix} 
    \sum_{m}h(\lambda_{m})V_{m,1}^{}V^{*}_{m,1} & * & \hdots & * \\
    * & \sum_{m}h(\lambda_{m})V_{m,2}^{}V_{m,2}^{*} & * & \vdots \\
    \vdots & * & \ddots & *\\
    * & \hdots & * & \sum_{m}h(\lambda_{m})V_{m,n_0}^{}V_{m,n_0}^{*}
    \end{bmatrix}  
\end{align}
where $\vec V_m$ is an eigenvector of $F$ associated with eigenvalue $\lambda_m$;
$V_{m,l}$ is its $l$-th component. 
We have denoted the number of electrons by $n_{0} =\lfloor L_{x}L_{y}/2\pi \rfloor $.
Thus, the EE per mode $\tilde S_A(k_l)$ is given by the $l$-th diagonal element of the matrix in Eq.~(\ref{mat1}).
It is plotted in Fig.~\ref{entropie_par_mode}.  

\section{Convergence}
\subsection{Convergence of Fourier expansion} \label{Convergece avec N} 
\begin{figure}
\centering
\includegraphics[width=\textwidth*2/3]{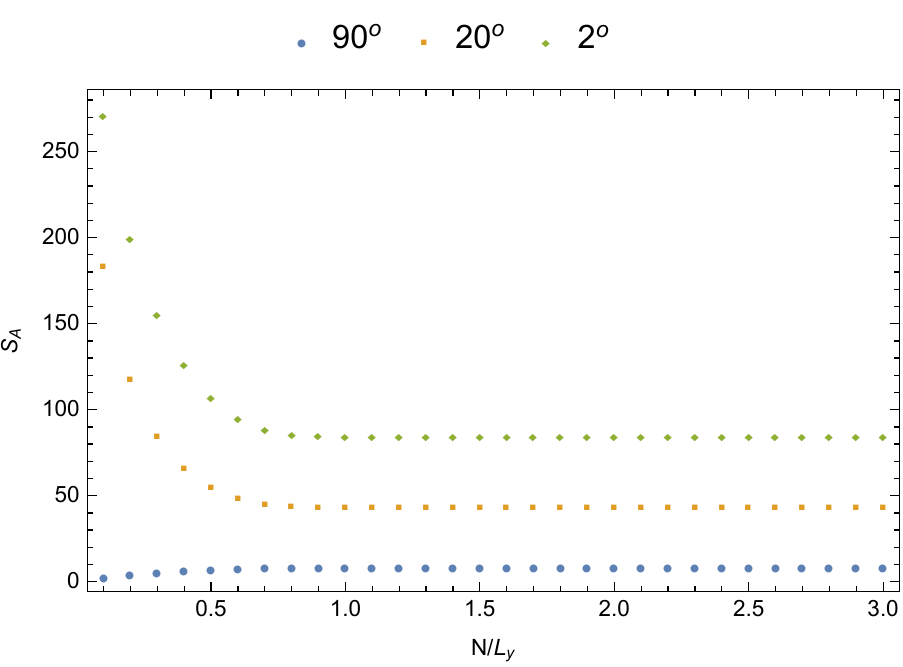}
\caption{Plot of $S_{A}$ as a function of $N/L_{y}$ for $\theta = 90^\circ, 20^\circ, 2^\circ$. Calculation performed with the following parameters : $L_y = 6\pi$, $L_x = 200$, $l_A = 40$}\label{entropie_convergence}
\end{figure}

Because we are using numerical approximations of the wavefunctions by means of a truncated Fourier decomposition,
we need to ensure that the number of Fourier modes used, $N$, is sufficiently large. 
In Fig.~\ref{entropie_convergence}, we show the EE as a function of $N$ for different values of $\theta$.
For all tested angles, the error becomes negligible when the number of modes roughly equals $2 L_{y}$ for values of $L_{y}$ smaller than $6\pi$.

To further confirm the convergence, the difference between our Fourier expansion and the exact solution was computed, which yielded a difference less than $10^{-9}$
for every point of the wavefunction, independently of the wavevector $k_{l}$.

\subsection{Convergence with $L_x$}\label{Convergece avec Lx}
The chiral modes induced by the physical boundaries lead to larger values of $L_{x}$ needed to ensure convergence compared to the torus geometry,
as shown in Fig.~\ref{fig_compar}. 
Furthermore, increasing $L_{x}$ produces oscillations of the EE with a period of $2\pi/L_{y}$ as discussed in Section~\ref{edge_modes}. 
Thus, if one wants to extrapolate the value of the EE to $L_{x}$ $\rightarrow$ $\infty$, it is simplest to use values of $L_{x}$ with spacing that corresponds to a multiple of the period. The EE obtained in the limit $L_{x} \rightarrow \infty$ should be independent of the starting point (one could decide to start at any point in a cycle), as we have numerically confirmed. 
\begin{table}[h]
    \centering
    \begin{tabular}{l|llll}
      & Estimate & Standard Error \\
    \hline
     $\zeta$ & $-1.2267$ & $3\times 10^{-4}$ \\
     $\bar S$ & $4.825240$ & $2\times 10^{-6}$ \\
    \end{tabular}
    \smallskip
    \caption{Value and error of parameters in the function used for the fit of the entanglement entropy with $L_{y} = 4\pi$, $l_{A} = 20$.}
    \label{convLx}
\end{table}

\begin{figure}
\centering
  \includegraphics[width=\textwidth*2/3]{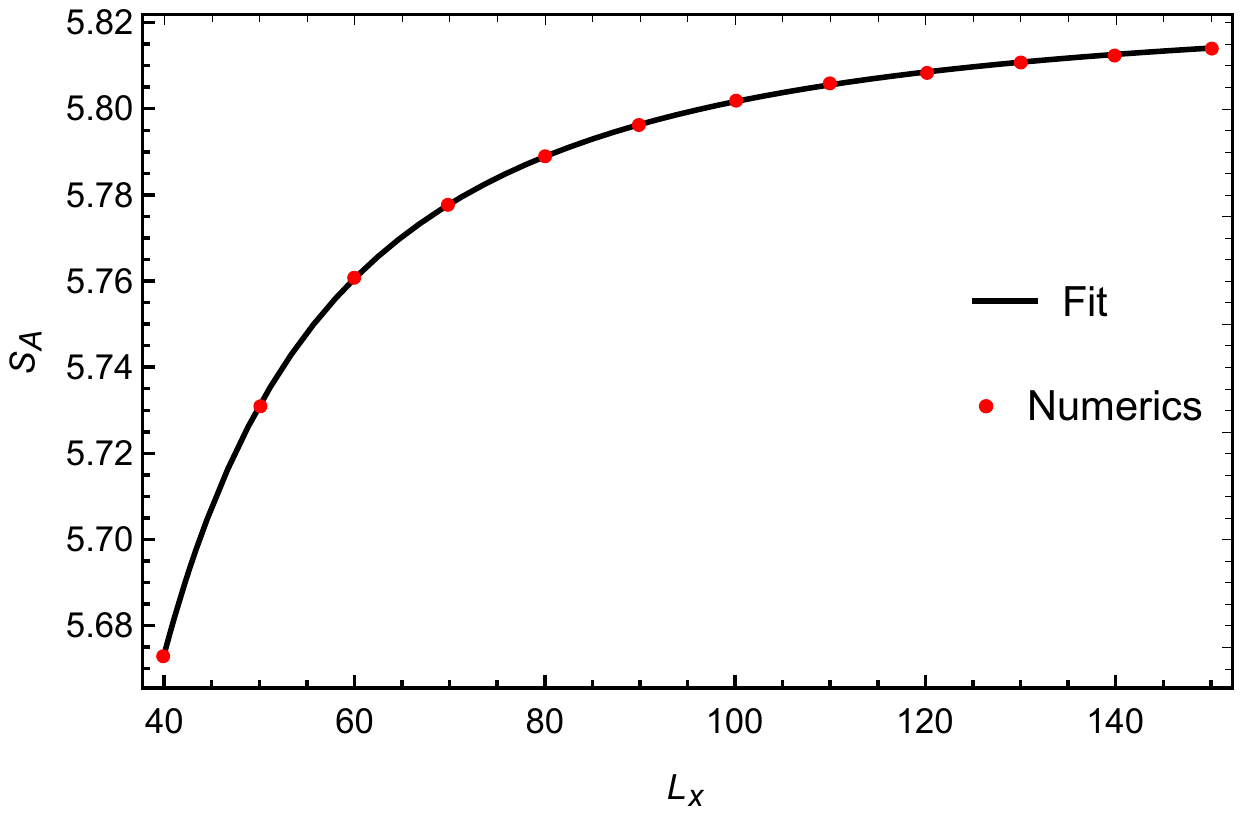}
  \caption{Fit of entanglement entropy as a function of $L_{x}$ with $L_{y} = 4\pi$, $l_{A} = 20$, $\theta$ = $\pi/2$.}
  \label{extrapolation_Lx}
\end{figure}

In order to perform the $L_x\to\infty$ extrapolation, we have used the function:
\begin{align}
  \frac{1}{3}\ln{\left(\frac{L_{x}}{\pi \ell_{B}}\sin{(\frac{\pi l_{A}}{L_{x}})}\right)} +\frac{\zeta}{L_{x}} + \bar S
\end{align}
The first term is the EE of a 1d CFT describing the edge mode contribution (with $\ell_B$ reinstated). We then added a $\zeta/L_x$ term. $\bar S$ is the $L_x=\infty$ asymptotic EE for given $L_y,\theta$.
An example of the corresponding fit is shown on Fig.~\ref{extrapolation_Lx}. 

\subsection{Small angles}\label{Convergece avec lb}

All the results presented in this paper are given for a region $A$ (see Fig.~\ref{illustration_boundary}) for which $l_{A}$, $L_{x}$ and $L_{y}$ are much larger
than the magnetic length $\ell_{B}$, which was fixed to 1 in this paper.
If the dimensions of $A$ were to be smaller than the magnetic length, our prescription for the the area law would not hold.
Now, when the angle $\theta$ becomes very small, the edges of the parallelogram $A$ could become too close to each other.
We found that imposing $l_{A}\tan(\theta) \gg \ell_{B}$ ensures that the EE behaves as expected. 
\begin{figure}
\centering
  \begin{subfigure}[b]{0.4\textwidth}
          \includegraphics[width=\textwidth*1]{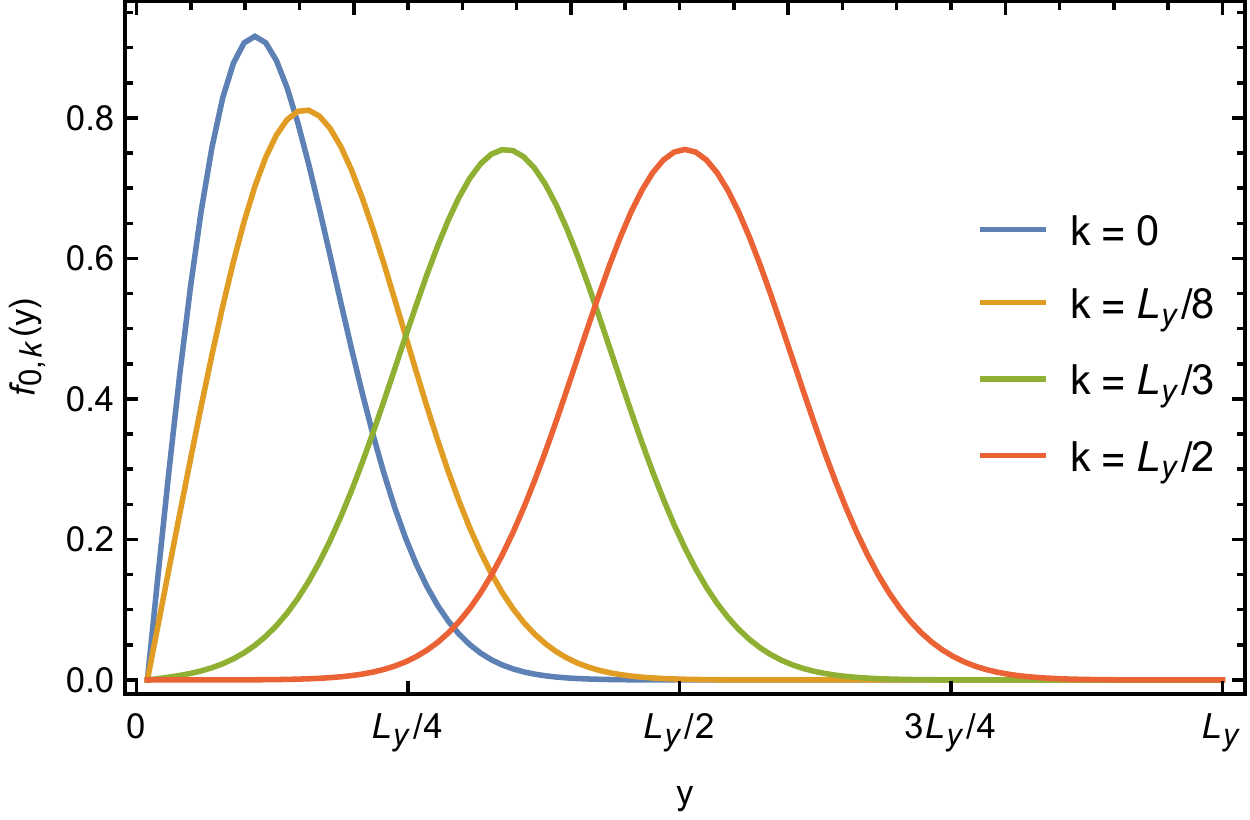}
    \label{fonction_sinus}
  \end{subfigure}
  \begin{subfigure}[b]{0.4\textwidth}
      \includegraphics[width=\textwidth*1]{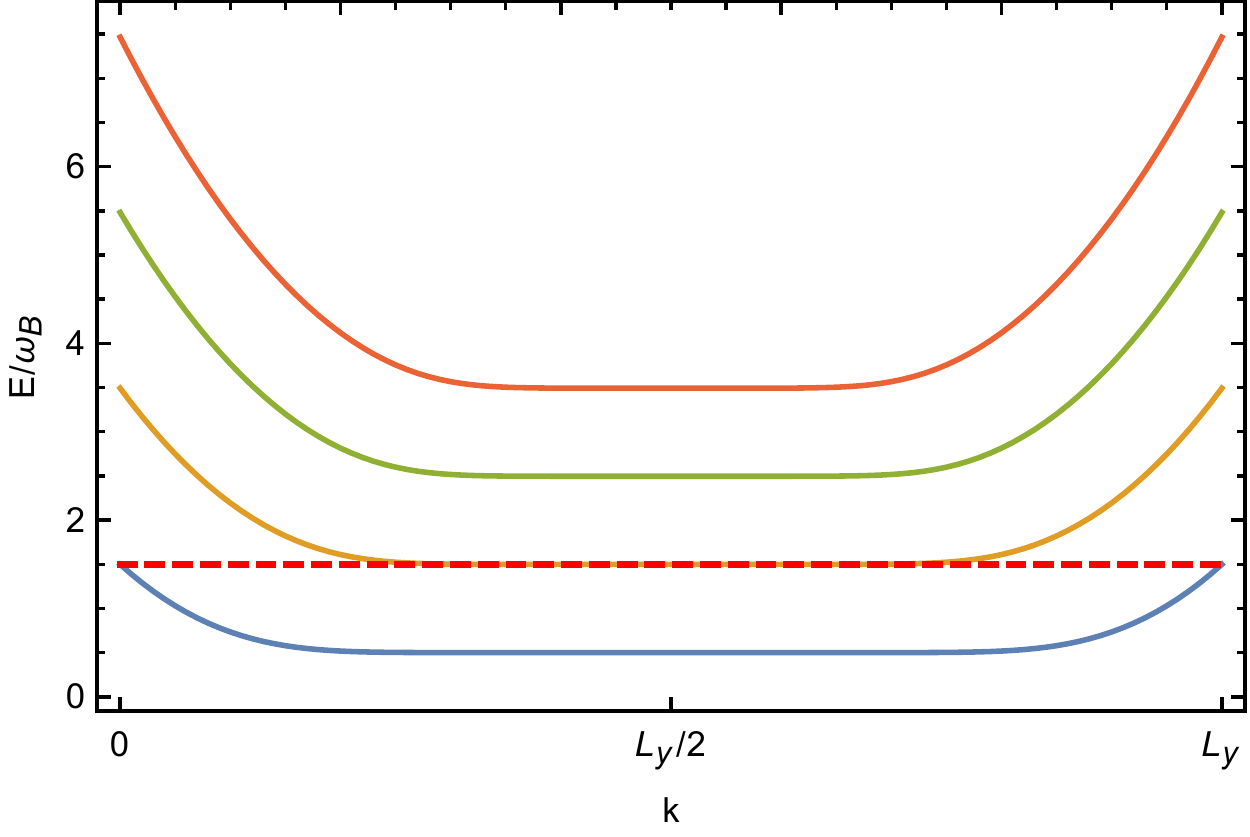}
    \label{fig:2}
  \end{subfigure}
  \caption{The left panel shows the wavefunction $f_{0,k}(y)$, $y\in [0,L_y = 10]$ for different values of $k$.
    The right panel shows the energy of the first 4 Landau levels normalized by $\omega_B = eB/m_e$.
    The red dashed line indicates the energy at the edges of the LLL which is the same as the energy of the plateau of $n=1$ LL.}
\label{fonction_sinus} 
\end{figure}  
 
\section{Exact solution}\label{Solutions_exactes}
Solving exactly the Schr\"odinger equation (\ref{y-eq}) yields the following wavefunction in the $y$-direction (up to an overall normalization): 
\begin{equation}
f_{n,k}(y) = D_{\frac{-E_{n} -1}{2}}\left(i \sqrt{2} (y-k)\right)-\frac{D_{\frac{-E_{n} -1}{2} }\left(-i \sqrt{2} k\right) D_{\frac{E_{n} -1}{2}}\left(\sqrt{2} (y-k)\right)}{D_{\frac{E_{n} -1}{2}}\left(-\sqrt{2} k\right)} 
\end{equation}
with 
\begin{equation} \label{transcendante}
D_{\frac{E_n -1}{2}} \left(-\sqrt2 k\right) D_{\frac{-E_n -1}{2}}
\left( i \left(\sqrt 2 L_y -\sqrt2 k\right)\right)-D_{\frac{-E_{n} -1}{2} }\left(-i \sqrt{2} k\right) D_{\frac{E_{n} -1}{2}}\left(\sqrt{2} L_{y}-\sqrt{2} k\right)=0
\end{equation}
The energy $E_n(k)$ is obtained by numerically solving Eq.~(\ref{transcendante}) for a given wavevector $k$. 
The function $D_{\nu}(z)$ is called the parabolic cylinder function and can be related to the hypergeometric function $_1F_1$:
\begin{align}
  D_{\nu}(z) = \frac{1}{\sqrt{\pi}}2^{\nu/2}e^{-z^2/4}\left(\cos{\left(\frac{\pi \nu}{2}\right)}\Gamma{\left(\frac{\nu +1}{2}\right)}
  {}_1F_{1}\left(\frac{-\nu}{2};\frac{1}{2};\frac{z^2}{2}\right) + \sqrt{2}z\sin{\left(\frac{\pi \nu}{2}\right)}\Gamma\left(\frac{\nu}{2} +1\right)
  {}_1F_{1}\left(\frac{1}{2} - \frac{\nu}{2};\frac{3}{2};\frac{z^2}{2}\right)\right)
\end{align}
with
\begin{align}
  _{1}F_{1}(a;b;z) = \sum_{n=0}^{\infty}\frac{(a)_{n}}{(b)_{n}}\frac{z^n}{n!}; \quad (x)_{n} = x(x+1)(x+2)\cdots(x+n-1)
\end{align}
The solutions are shown in Fig.~\ref{fonction_sinus}.

\newpage
\bibliography{bibliographie.bib}
\end{document}